# Electrostatics controls the formation of amyloid-like superstructures in protein aggregation

*Vito Foderà[1,] *, Alessio Zaccone[1,2,] *, Marco Lattuada[3,†] and Athene M. Donald[1]*

[1]Sector of Biological and Soft Systems, Department of Physics, Cavendish Laboratory, University of Cambridge, JJ Thomson Avenue, Cambridge CB3 0HE, United Kingdom.

[2]Theory of Condensed Matter, Department of Physics, Cavendish Laboratory, University of Cambridge, JJ Thomson Avenue, Cambridge CB3 0HE, United Kingdom.

[3] ETH Institute for Chemical and Bioengineering, HCI F135, Wolfgang Pauli Strasse 10, 8093 Zurich, Switzerland.

**CORRESPONDING AUTHOR FOOTNOTE:** *To whom correspondence should be addressed. Vito Foderà: vf234@cam.ac.uk; Alessio Zaccone: az302@cam.ac.uk

[†] Present address: Adolphe Merkle Institute University of Fribourg  Route de l'ancienne Papeterie CP 209 CH-1723 Marly 1, Switzerland

# Abstract

The possibility for proteins to aggregate in different superstructures, i.e. large-scale polymorphism, has been widely observed, but an understanding of the physico-chemical mechanisms behind it is still out of reach. Here we present a theoretical model for the description of a generic aggregate formed from an ensemble of charged proteins. The model predicts the formation of multi-fractal structures with the geometry of the growth determined by the electrostatic interactions between single proteins. The model predictions are successfully verified in comparison with experimental curves for aggregate growth. The model is general and is able to predict aggregate morphologies occurring both *in vivo* and *in vitro*. Our findings provide for the first time a unifying and general framework where the physical interactions between single proteins, the aggregate morphology and the growth kinetics are connected into a single model in excellent agreement with the experimental data.

Understanding the connection between growth mechanisms and morphology is a central problem for modelling self-assembling biological systems[1]. This basic topic in condensed matter and biophysics was already emphasized by the far-seeing work of D'Arcy Thompson at the beginning of the last century, focusing on the need for quantitatively describing the specific physical interactions leading to different structural arrangements[2]. Protein aggregation is a central area in current biophysics research and the field, directly and with equal importance, spans both basic and applied sciences[3]. In fact, investigating the origin of neurodegenerative diseases like Alzheimer's and Parkinson's cannot be separated from an accurate description of the inter- and intra-molecular interactions involved in the formation of such protein aggregates[3]. An increasing interest has recently been addressed towards understanding the occurrence of pronounced microscopic polymorphism in the formation of mature aggregates and, specifically, of aggregate of amyloid origin, i.e. elongated fibrils[4-6]. However, together with this structural polymorphism of fibrils, it has been widely observed that, both *in vivo* and *in vitro*, amyloid aggregates may generally conserve their basic structural arrangement of cross β-sheet, yet exhibit significantly different packing into three dimensional superstructures (µm range), i.e. macroscopic or large-scale polymorphism. However, the mechanisms underlying the occurrence of a particular superstructure are still unknown.

Under destabilizing conditions, a number of model globular proteins have been shown to aggregate into different forms, mainly depending on the pH of the solution,[7] or generally on the charge state affecting their stability[8] and shear fileds[9]. Close to the isoelectric point (pI) of the protein (i.e. where there is no net charge on the protein), compact spherical aggregates with radius up to 1 µm (particulates) are detected[10]. On the other hand at low pH far from the pI (i.e. high charge on the protein), elongated amyloid fibrils[11] occur together with a fascinating morphology known in the literature as an amyloid spherulite[12]. Spherulites (with radii up to hundreds of µm, Figure S1 in Supplementary Information, SI) are thought to be composed of a central and compact part (referred to as the *precursor*) surrounded by a low-density outer part (often referred to as the *shell*)[13]. They are rich in β-structures[12], show a positive labelling when bound to amyloid sensitive dyes[12] and recently their possible role in human amyloid pathologies has been also reported[14]. However, even though the occurrence of such a variety of morphologies is widely observed, several basic questions remain unanswered: what is the connection between the macroscopic final morphology and details of the growth kinetics? How and why does the aggregate density change during growth? These questions lead also to the central and still unexplored problem of linking the aggregation kinetics curves with the microstructural details of the growing aggregate. Several models based on nucleation and aimed at rationalizing the aggregation process assume generic mechanisms for protein assembly and disassembly[15]. However, these models do not allow for a proper connection between the large-scale morphology of the aggregates and the interactions and phenomena happening at shorter time and length scales.

Here we bridge the gap between the kinetic description of the overall process, the predictions of large-scale morphologies and the microphysics of protein interactions. We show that a generic aggregating ensemble of charged proteins intrinsically evolves through a multi-fractal growth. By means of a microscopic thermodynamic model, we describe the dependence of such growth upon the electrostatic interactions between single protein molecules. In the case of spherulite-forming reactions, we implement the microscopic model and numerically solve the master kinetic (population balance) equations for the kinetics. By comparison with experimental data, we prove that the model quantitatively predicts both the overall kinetics and the large-scale morphology on the basis of the microphysics. Importantly, we also show that the proposed framework is general and can be used to recover the amyloid fibril morphology in the limit of high protein charge and the occurrence of particulates in the limit of uncharged proteins, thus providing a unifying framework for the description of the protein aggregation kinetics.

## RESULTS

### Theoretical model for a growing aggregate

We base our model for aggregate formation on the calculation of the free energy of a spherical cluster with radius $R$:

$$\Delta F = -\frac{qU}{2} + \frac{3\gamma}{R}\frac{4\pi R^3}{3} + \frac{(N \cdot n_c \cdot e)^2}{8\pi\varepsilon\varepsilon_0 R} - kT \cdot \log(X_N / N) \qquad (1)$$

where $q$ is the total number of contacts between pairs of molecules in the aggregate, $U$ is the energy per contact, $\gamma$ is the surface tension[15], $R$ is the radius of the cluster, $N$ the number of the molecules in the cluster, $n_c$ is the effective number of charges on a single particle, $e$ is the elementary charge, $\varepsilon$ and $\varepsilon_0$ are the relative dielectric constant and the permittivity in vacuum and $X_N$ is the volume fraction of molecules in the cluster. The first term is related to the binding between single molecules in a cluster and the second one is the correction factor due to missing contacts at the surface[12]. The third term represents the electrostatic (Born) energy required to move a charged protein from infinity to the charged aggregate[16, 17]. This term is related to the overall electrostatic repulsion between the molecules[16, 17] and, due to the complexity of the electrostatic interaction between two proteins, $n_c$ cannot be referred only to the absolute number of charges on the protein, but depends on several other factors. Proteins contain polar groups that are far from being evenly distributed so that the protein surface is not uniformly charged, generating large electric dipole and multipole moments that strongly affect the inter-particle interactions[18]. Moreover, contributions due to the water structuring around the protein[19, 20], hydrophobic interactions[20] and free electrolyte mediating the interaction between proteins[21] can further contribute to the overall electrostatic interactions between two molecules. For these reasons, the electrostatic term in equation 1 represents an effective term including all the above mentioned contributions. Finally, the last term in the equation 1 is the entropic contribution arising from the loss of translational degrees of freedom when particles are bound to the cluster[22].

Equation 1 can be rewritten (see SI for details) for the case of fractal growth to obtain the free energy of the cluster as a function of $R$, the fractal dimension $d_f$, the radius of the single protein $a$, the number of particle nearest-neighbors $Z$, the effective number of interactions $f$ and a binding energy $n_E$ kT.

$$\Delta F = -\frac{Z}{2}\left(\frac{R}{a}\right)^{d_f} n_E kT + \frac{f \cdot n_E kT \cdot d_f}{a^{(d_f-1)}} R^{(d_f-1)} + \frac{n_c^2 \cdot e^2}{8\pi\varepsilon_0 a^{2d_f}} R^{(2d_f-1)} - kT \cdot \log\frac{a^3}{R^3} \quad (2)$$

Equation 2 represents the free energy for an aggregate growing with a generic $d_f$. We evaluate eq. 2 for a spherical growing aggregate with $d_f = 3$ and for single (globular) proteins of radius $a = 2\ nm$ with an effective number of charges $n_c = 0.5$ (see details in the SI). We assume a binding energy of 10kT which is compatible with growth with $d_f \leq 3$ [16]. Importantly, in the specific case of aggregating proteins, the first member of the right hand side of eq. 2 refers to the binding energy between already destabilized and aggregation prone molecules. Figure 1a shows the free energy profile as a function of the cluster radius for spherical aggregates. The free energy of clustering shows an initial constant value and then a minimum of ~ -4.5x10$^{-16}$ J occurs at R ~ 40 nm. Above this radius value the function steeply and indefinitely increase towards positive values (Figure 1a). Varying the effective number of charges $n_c$ in eq. 2 affects the free energy profile (Figure 1b). Specifically, when $n_c$ is increased, the depth of the minimum decreases. Moreover, the position of the energy minimum shows a well-defined exponential decay as $n_c$ increases (inset in Figure 1b).

**Electrostatics controls the formation of superstructures**

Data in Figure 1 predict a growth of a spherical aggregate with the size dependence controlled by the electrostatic term in eq. 2. After reaching the energy minimum, the aggregate can no longer evolve with the same geometric features. In Equation 2 information about the structure is encompassed by the fractal dimension $d_f$. Calculations of the energy profile have been performed at 5 different $d_f$ values down to $d_f =1.75$. When the $d_f$ is decreased, the free energy minimum turns out to be shifted towards higher values of radius. This means that, after reaching the first minimum ($d_f =3$), the aggregating system can explore new minima of its free energy only if the morphology of the growth changes, i.e. if $d_f$ decreases. Since this change happens continuously, this leads to a multi-fractal profile for the free energy as shown in Figure 2a for 5 discrete values of $d_f$. The system will follow a pathway of energy minimization (in red in Figure 2a) leading to an aggregate with a compact central structure with $d_f =3$ (hereafter called *precursor*) and an outer part with a decreasing fractal dimension as a function of the radius (hereafter called *shell*). Importantly, the decrease in $d_f$ after the precursor formation proceeds continuously rather than in discrete steps. The model proposed here allows one to calculate how the multi-fractal profile evolves during the aggregate growth at different values of $n_c$. After the precursor formation, $d_f$ shows an exponential decrease for $n_c=0.5$ and $n_c=1$ (Figure 2b). Interestingly, strongly decreasing the $n_c$ value down to 0.001 (i.e. almost no net charge on the molecules) leads to an aggregate growing with $d_f =3$ for tens of µm (Figure 2b) before a significant decrease in $d_f$ can take place.

Extrapolating the $d_f$ *vs* R relationship from Figure 2b also allows us to quantitatively estimate the change in density during the aggregate growth compared to the precursor (Figure 2c, SI for details). After the formation of the precursor, a decrease of the density is predicted with the shell reaching a density $10^4$ times smaller than the precursor within 1 µm of growth for the data at $n_c=0.5$ and $n_c=1$. For the data at $n_c=0.001$, a significant decrease of the density is expected only when the aggregate reaches a radius > 20 µm (in blue in Figure 2c). In the limit of $n_c= 0$, $d_f$ is constant and equal to 3 for the entire growth, i.e. particulates. Conversely, for $n_c > 2$, the growth basically proceeds with $1< d_f < 2$ from the early stages, i.e. elongated fiber-like structures (Figure S2 in SI). The predictions of our model clearly explain the range of aggregate superstructures observed both *in vivo* and *in vitro* for several protein systems, i.e. particulates (pI of the protein) and amyloid fibrils and spherulites (far from the pI of the protein). They are sketched in Figure 2d. It is worth noting that the $n_c$ values used in the calculations account for the average effective number of charges on a single protein. However, it is well known that the dielectric properties of a protein are not constant along its length[23]. This means that electrostatic interactions of very different extent (i.e. with different $n_c$ values) could take place within the same ensemble of charged proteins, determining the simultaneous occurrence of different morphologies. This is indeed experimentally verified for a number of protein systems that, in specific conditions, show the presence of both spherical and elongated aggregates[13, 24].

**Prediction of the experimental growth curves**

Now the question is if we can quantitatively describe the temporal course of experimentally observed aggregate growth by our microscopic model. To answer this question, we focus on the growth of spherulites and consider the recently published static light scattering data for spherulite growth in samples of bovine insulin during incubation at 60°C and at different pH values in the range 1-1.75 [25]. Decreasing the pH in this range would mainly increase the positive charge on the protein, changing the electrostatic interactions between the molecules[26].

Aggregation kinetics (symbols) in Figure 3a show the well-known sigmoidal profile, with an initial lag time which decreases as the pH is lowered. Importantly, in these experimental conditions (see Methods) and at the end of the aggregation kinetics, native insulin molecules are mainly converted into amyloid spherulites[12, 25], so that these data are representative of spherulite growth and hence suitable for a comparison against our theoretical model. A closer view to the lag time shows an increase in the signal already in the very early stages of the process for all the investigated conditions (Figure 3b). After that, an abrupt increase in the growth rate of aggregates characterizes the temporal profile before reaching a plateau. To date, no quantitative description has been reported for such a two-step increase and only generic and qualitative explanations are suggested for the early increase of the signal[27, 28]. In order to compare the predictions of our theoretical framework with experimental data we consider the master kinetic (population balance) equations for the aggregation process

$$\frac{dC_k}{dt} = \frac{1}{2}\sum_{i+j=k} K_{ij} C_i C_j - C_k \sum_{i=1}^{\infty} K_{ik} C_i - K_k^B C_k + \sum_{i=k+1}^{\infty} K_{ik}^B C_i \qquad (3)$$

where $C_i$ is the concentration of aggregates with mass $i$ (i.e., made of $i$ protein molecules), and $K_{ij}$ is the kernel determining the rate of aggregation between two aggregates, one with mass $i$ and the other with mass $j$ (see SI for full details). The last two terms accounts for thermal breakup of a cluster of size $k$ and generation of a $k$ cluster by breakup of a cluster of size $k+i$. For systems in which the thermal breakup is not relevant the last two terms are negligible. This is actually the case of our system (~10kT, see section 3 in SI for details) so that we need to solve the following equation:

$$\frac{dC_k}{dt} = \frac{1}{2}\sum_{i+j=k} K_{ij} C_i C_j - C_k \sum_{i=1}^{\infty} K_{ik} C_i \qquad (4)$$

The microscopic rates can be calculated based on a conventional diffusion-limited aggregation scheme (see SI for details) and they fully account for four basic interactions, which have all been computed using the Derjaguin-Landau-Verwey- Overbeek (DLVO) theory: 1) monomer-monomer,

2) oligomer-oligomer, 3) shell-monomer and 4) shell-oligomer interaction. The geometry of the growth is also taken into account by implementing the $d_f$ evolution predicted by the theory into the master equation (see SI for details). Together with the above hypotheses, pairs of values of $n_c$ and precursor radius, as obtained from the model (inset in Figure 1b and Tab. S1 in SI) have been used to simulate curves with different electrostatic properties. Simulations (solid lines in Figure 3a) are able to predict both the initial slow increase in the light scattering curves (solid lines, Figure 3b) and the rapid growth of signal before reaching the plateau. The good agreement can also be seen by considering the experimental lag times versus the theoretical prediction (Figure 3c).

**On the growth and form**

We can now go back to our original question: can we relate the temporal curve with details of the large-scale morphology of the growing aggregate? Electrostatic potentials between all the species during the process and the temporal evolution for each species can be calculated (eq. 2.10 in SI). Figure 4a and 4b show the potential curves for the interactions considered in the process and changes of oligomer and precursor populations as a function of time, respectively. Moreover, in Figure 4b the scattering curve (dashed purple line) is also shown to visualize the profile of the multifractal growth. All these data are for the kinetics at pH 1; analogous trends were obtained at other pHs. The association of individual proteins proceeds without any significant energy barrier (black line in Figure 4a), so that a rapid formation of oligomers takes place in the early stages of the process. As a consequence, already after a few minutes of incubation, our simulations predict an almost complete depletion of monomers leading to the formation of spherical oligomers with an average radius of ~16 nm (red circles in Figure 4b). This is in agreement with previous experimental data on insulin amyloid aggregation showing the presence of oligomers of comparable size as the one predicted by the model.[29, 30] Such aggregates appear to be stable during the entire duration of the lag phase (Figure 4b). Moreover, the formation and presence of such species already in the early stages of the process explain the initial increase observed in the light scattering curves (Figure 3b). Afterwards, the kinetics is entirely controlled by the interaction between oligomers. Oligomer association represents the bottleneck for the activation of the precursor-shell growth: they associate (decrease of the oligomer fraction, red circles in Figure 4b) until they reach a critical radius leading to a specific potential barrier for oligomer-oligomer interaction up to $288 \times 10^{-21}$ J (~ 70kT). This barrier makes further association between oligomers with critical radius extremely unlikely (red line in Figure 4a). This critical size defines the radius of the precursor, the number of which increases until the end of the lag phase (~ 4000 s, blue triangles in Figure 4b); after that shell growth is dominant and takes place through association between precursors with smaller oligomers and/or residual monomers. This shell growth can proceed without any significant barrier (blue line in Figure 4a), leading to the consumption of the precursor population (4000-10000 s, blue triangles in Figure 4b) and the formation of the multi-fractal structure (abrupt increase in the scattering curve, dashed purple line in Figure 4b). These quantitative results are sketched in Figure 4c where monomers, oligomers and precursors are represented by green, red and blue spheres, respectively.

Our model suggests that the difference in the lag time of the kinetics in Figure 3a is basically related to the radius of the precursor. The precursor radius as calculated by the simulations increases by ~10% passing from pH 1 to pH 1.75: when the precursor is smaller, the time necessary to reach the critical radius for the shell growth is reduced. Such evidence is well summarized by the linear correlation between the experimental lag time and the precursor radius (Figure 4d). After the precursor formation and depending on the specific electrostatic interactions between the particles (Figure 2), the shell starts growing, producing the abrupt growth of the aggregate size leading to the peculiar structure of the spherulites (Figure 2d). It is worth noting that for a number of amyloidogenic systems the quick growth has been elegantly explained in terms of secondary nucleation processes, mainly due to fragments of amyloid fibrils that act as nucleation points for new fibrils[15] and autocatalytic effect at the fibril surfaces[30-32]. However, in the experimental conditions investigated here, the formation of free amyloid fibrils is not the main pathway[24] and the kinetics curves should be considered as representative of spherulite growth. This gives rise to the need to know the morphology of the growing aggregate to effectively predict the molecular mechanism involved in the secondary nucleation process. In the specific case of systems mainly forming spherulites, the speeding up of the process is basically dictated by the change in growth geometry from a compact sphere (precursor) to an increasingly less compact geometry (shell), leading to the reported hedgehog-like structure[12]. The need to minimize the free energy of clustering (which is dominated by the electrostatic contribution) induces the system to grow faster than the available volume due to multi-fractality setting in. In turn, this multi-fractality causes the observed loss of compactness and the associated rapid growth in the intensity of scattered light. This mechanism can be compatible with the classical heterogeneous nucleation/growth on aggregate surface, i.e. growth of low-fractality structures on the surfaces of compact spherical aggregates (precursor). Importantly, our mathematical framework can recover the classical nucleation theory in the limit of weak attraction (please see section 3 in SI).

**Discussion**

Using a combination of theoretical arguments, quantitative experiments and simulations, we show that multi-fractal patterns arise in protein aggregation reactions due to the interplay of a random multiplicative process (growth) which evolves under the constraint of following a path of minimal free energy, the latter being dominated by electrostatics. Our approach naturally explains the *in vivo* and *in vitro* occurrence of a range of protein aggregate structures, i.e. particulates, spherulites and amyloid fibrils, controlled by electrostatic interactions. Moreover, implementing the proposed theory in the master kinetic equation for the growth of spherulites gives an excellent quantitative agreement with experimental data.

In view of the absence of restrictive assumptions in the proposed model, our framework is not limited to proteins and equation 2 can be used to describe generic systems of charged particles undergoing random multiplicative and branching processes. This makes our approach broadly applicable to a variety of systems, both in physics and chemistry, where multi-fractal patterns arise[33]. A classic example is the dielectric breakdown of insulators where the electrical discharges propagate through the material following multi-fractal patterns. Interestingly, also in this case, electrostatics represents the microscopic cause of multi-fractality in that the discharge tree must propagate by minimizing the electrostatic energy density due to the charge carriers[34, 35]. The present work shows for the first time that similar pathways can also occur in such biologically relevant processes as protein aggregation, explaining the observed variety of amyloid-like superstructures. This insight is crucial also in a medical perspective. For instance, modulating protein-protein interactions by small molecules is nowadays the prominent route for designing potential inhibition strategies of the *in vivo* aggregation processes[36]. A successful design of effective inhibitors is obviously dependent on an adequate knowledge on how inter-protein interactions are related to both the overall aggregation kinetics and the aggregate morphology. Our framework provides the unprecedented possibility to connect these three aspects, offering a new tool to single out, rationalize and control the mechanisms behind protein aggregation phenomena.

**Methods**

*Sample preparation and light scattering measurements* Bovine Insulin (BPI) was obtained as a lyophilised powder from Sigma Aldrich (I5500). Solutions at protein concentration of 4 mg/ml were prepared dissolving the powder in water with 25 mM NaCl and aliquots of 10% v/v HCl were added to the solutions to reach the desired pH value. pH measurements have an accuracy of pH ± 0.01. Aggregation was thermally induced at 60°C. Details on sample preparation and light scattering measurements were previously reported[25].

*Population Balance Equation model.* Simulations based on population balance equations have been performed by solving equation 3 using the method proposed by Kumar-Ramkrishna[37]. All the details are reported in the Supplementary Information. The change in fractal dimension during growth has important implications on the calculations of the reaction rates in equation 3: oligomers with a size smaller than the precursor are treated as a spherical cluster of charged particles, and have aggregation rates decreasing as their size increases. On the other hand, in the outer shell formation, the interactions most likely involve one single monomer unit located on the outer surface of the low density shell. Their reactivity is thus assumed to be equal to that of monomers[38] and all the species in the model are allowed to interact to each other (see Figure S3 and related discussion in SI). Importantly, the simulations have been obtained by approximating the multi-fractal radial profile of the shell by means of a single $d_f= 1.9$ which gives the best one-parameter fit of the multi-fractal aggregate structure factor (see Figure S4 in SI). This approximated structure factor is the one that has been used to calculate the time evolution of the scattering intensity (such a task would be computationally prohibitive if we were to use the actual multi-fractal form factor).

**Acknowledgments**

We thank Mike Smith (University of Nottingham) for help in designing the experiments. We thank James Sharp and Clive Roberts (University of Nottingham) for useful discussions. V.F. thanks Lorenzo Di Michele (University of Cambridge) for useful discussions. V.F. thanks Bente Vestergaard (University of Copenhagen) for the inspiring discussions on polymorphism and superstructures in protein aggregation. Funding from the EPSRC (EP/H004939/1) and the Swiss National Science Foundation (Grants Nr. PBEZP2-131153 and Nr. 200020-126487/1) is gratefully acknowledged.

**Author contributions**

V.F., A.Z. and A.M.D. were responsible for work concept, planning and management. A.Z. developed the theoretical model in close collaboration with V.F. V.F. was responsible for the experimental part and for carrying out the analytical calculations of the model. M.L. was responsible for carrying out the numerical simulations. V.F., A.Z., M.L. and A.M.D. were responsible for interpreting the results. V.F. wrote the paper in collaboration with A.Z., A.M.D. and M.L.

# Figure Legends.

**FIGURE 1 Free energy landscape for a growing cluster.** (a) Free energy calculated by means of equation 2 for $d_f = 3$ considering an effective charge on a single particle ($n_c e$) of 0.5e. The profile shows an energy minimum at ~ 40 nm. The cluster can grow with a compact and spherical structure ($d_f = 3$) until it reaches the radius of 40 nm. (b) Energy minimum profiles ($d_f = 3$) at different values of the effective charge $n_c e$ (0.37e-1.5e). Inset: energy minimum position as a function of $n_c$. The model shows that the charge on a single particle determines the maximum radius reachable for a growing spherical cluster.

**FIGURE 2 Amyloid-like superstructures are controlled by electrostatics.** (a) Free energy profile for a growing cluster as a function of the radius calculated at 5 different fractal dimensions by means of equation 2 ($n_c e = 0.5e$). After reaching the first energy minimum ($d_f = 3$) the cluster can further growth only if its $d_f$ decreases. The solid red line indicates the most energetic favorable pathway for the aggregate growth. (b) Fractal dimension and (c) density of the aggregate normalized by the precursor density during the cluster growth: $n_c e = 0.001e$ (blue), $n_c e = 0.5e$ (green) and $n_c e = 1e$ (red). (d) Illustrative sketch of the morphologies: from compact spherical aggregate with constant density (in the limit of $n_c = 0$) i.e. particulates, to elongated structures ($n_c > 1$,), i.e. amyloid fibrils (Figure S2 in SI). For $0 < n_c < 1$ precursor-shell growth is predicted (i.e. spherulites).

**FIGURE 3 Prediction of the experimental growth curves** (a) Static light scattering intensity as a function of time during insulin spherulite formation at different pH. Increasing the pH implies a decrease of the effective charge on the protein molecule. Solid lines represent simulated curves according with the theoretical model. Values of the precursor radius and effective charge from the theoretical model are used in the calculation of equation 3 together with the hypothesis of multi-fractal growth. (b) Zoom on the early stages of the process. (c) Comparison between experimental and simulated lag times of the process. Error bars represent absolute deviations observed on three replicates. The loss of compactness of the aggregate during the growth is responsible for the rapid increase in the intensity of scattered light.

**FIGURE 4 Early stages and activation of the multifractal growth** (a) Energy potentials for monomer-monomer (dashed black), oligomer-oligomer (close to the critical size of the precursor, red) and shell-monomer/shell-oligomer interactions (dashed blue). With the word "oligomer" we refer to an aggregate with a number of units higher than 1 (monomer) and lower than the number of molecules in the precursor. Association between oligomers with critical radius (precursors) is extremely unlikely (red energy barrier) and the fractal dimension for the growth changes (shell). (b)

Oligomers and precursor population during the aggregation process as predicted by the model. The scattering curve is shown to visualize the activation of the multifractal growth (c) Sketch of the mechanisms during the lag time as predicted by the theoretical model and in agreement with the experimental kinetics curves: green (monomers), red (oligomers) and blue (precursor). (d) Experimental lag time as a function of the simulated precursor size; the linear relationship indicates that the lag time is the time required to form the precursor and activate the growth with $d_f < 3$.

**FIGURE 1**

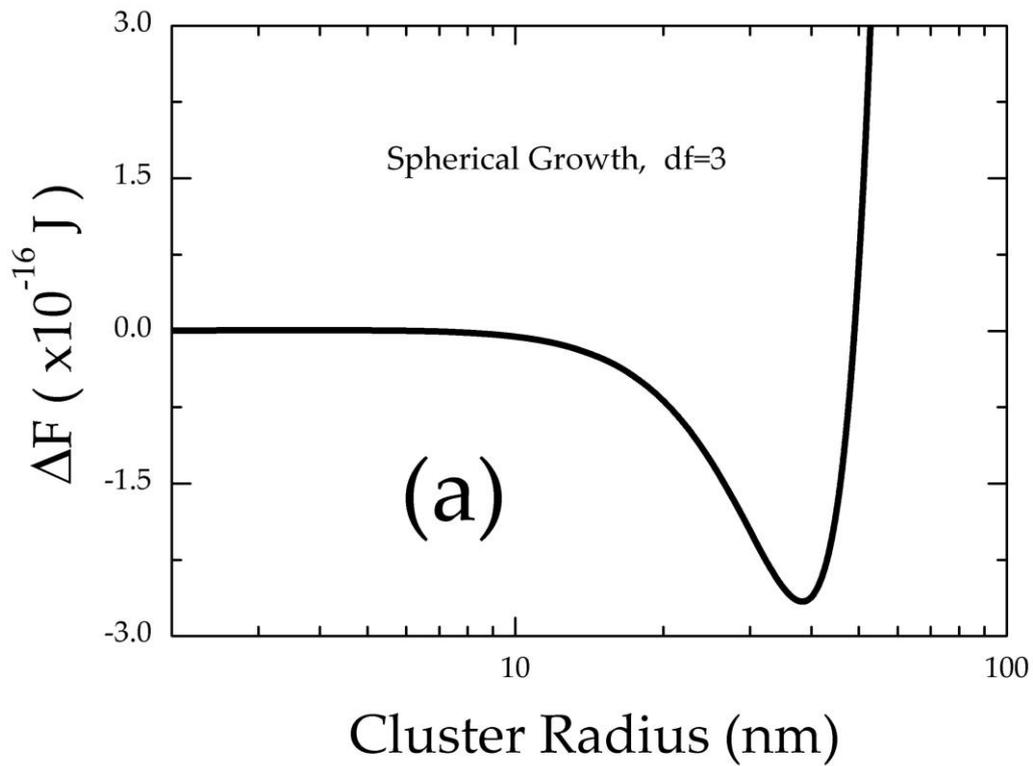

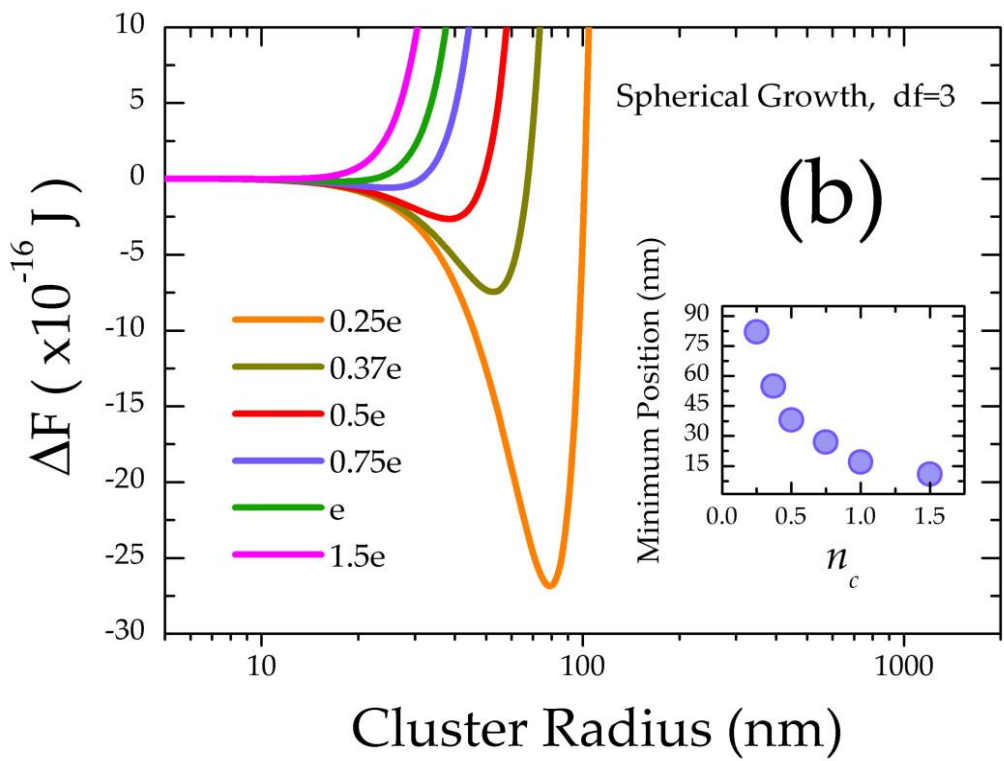

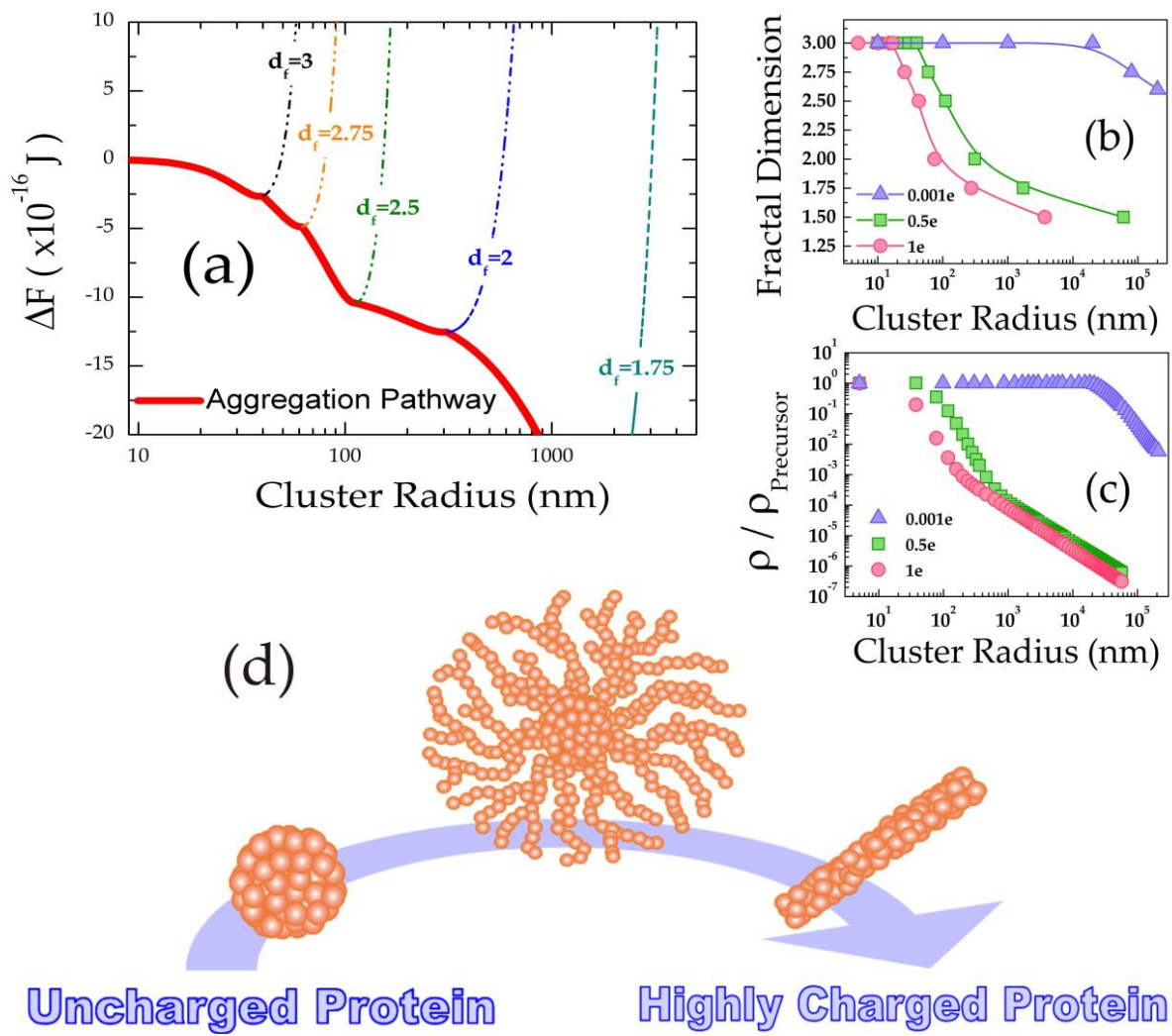

**FIGURE 3**

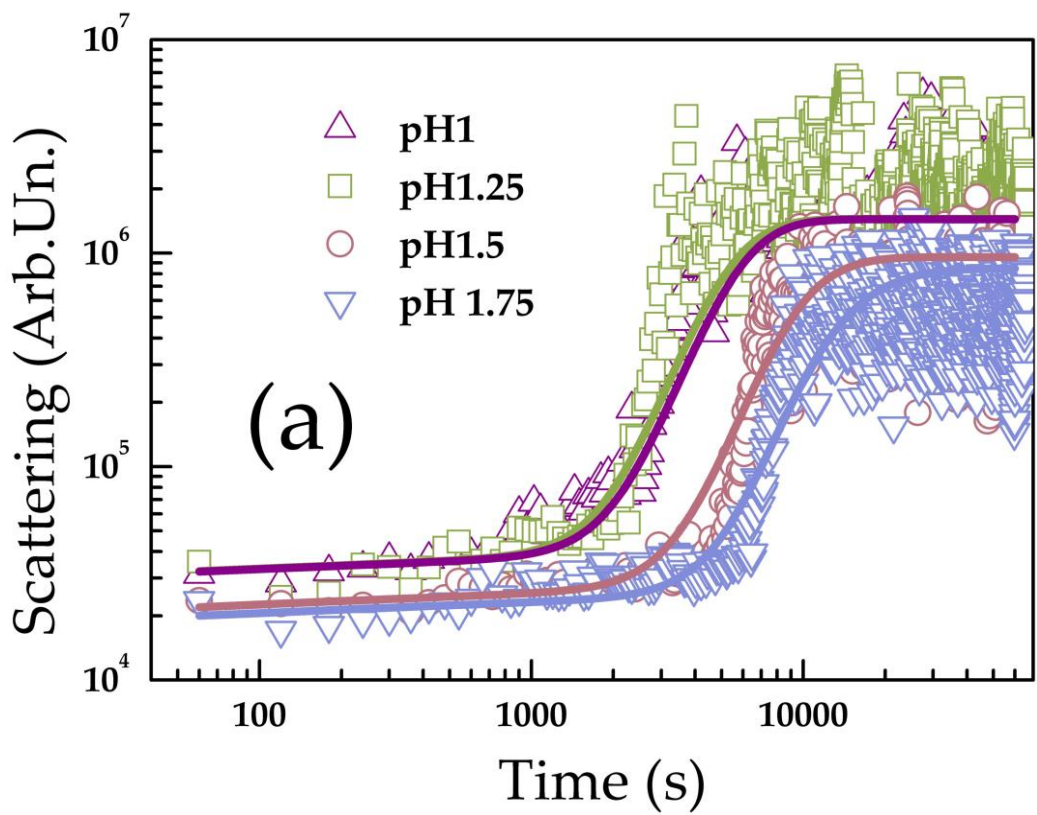
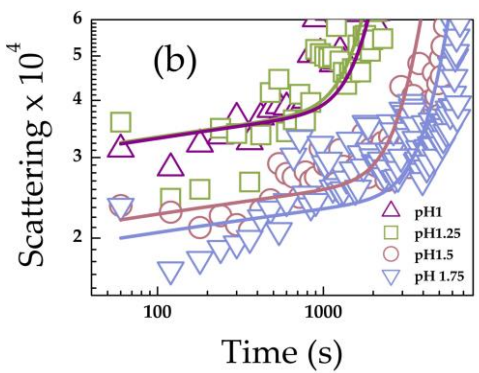
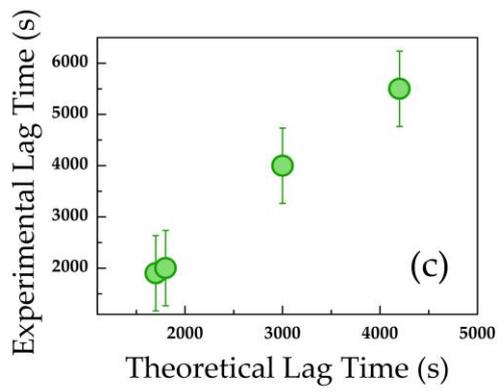

**FIGURE 4**

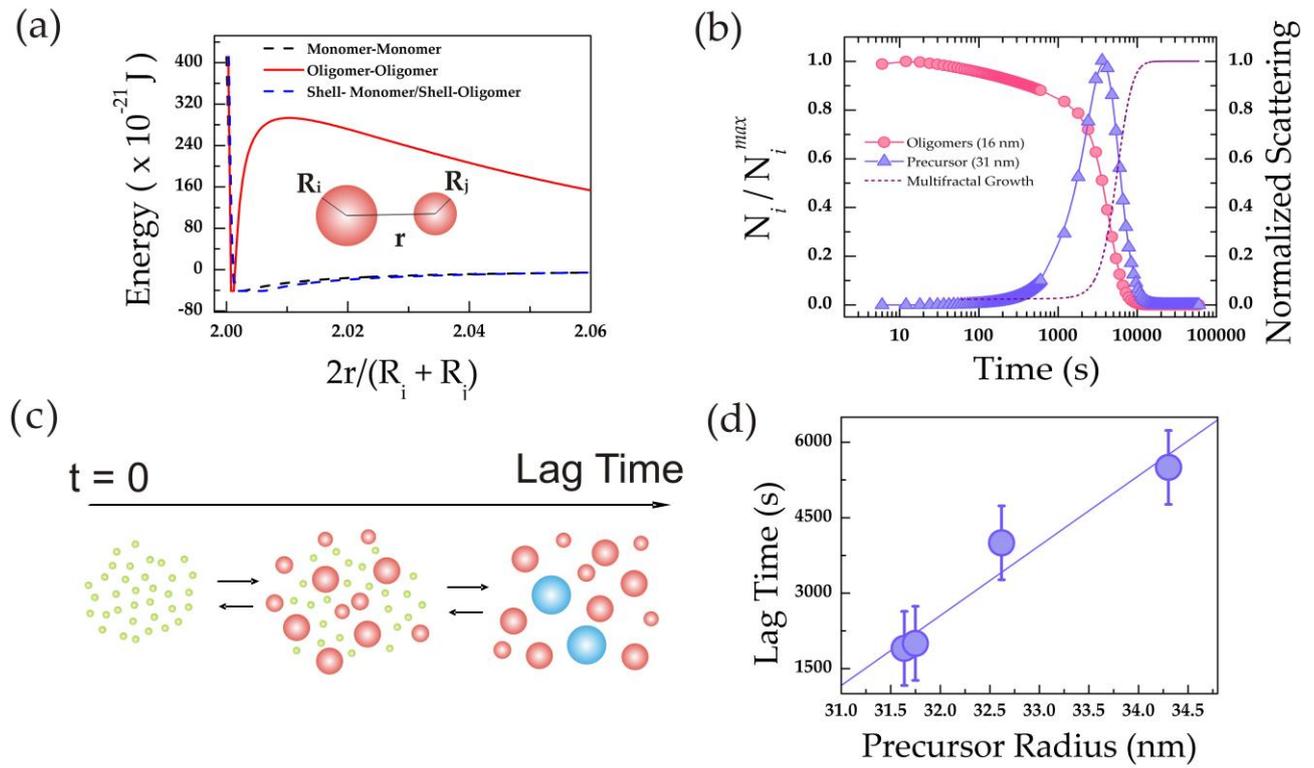

Electrostatics controls the formation of amyloid-like superstructures in protein aggregation

*Vito Foderà, Alessio Zaccone, Marco Lattuada, Athene M. Donald*

**Supplementary Methods**

**1 Free energy of a growing aggregate.**

*Energy calculation*

Consider the free energy of a spherical cluster with radius R with respect to the single particle forming the cluster:

$$\Delta F = -\frac{qU}{2} + \frac{3\gamma}{R}\frac{4\pi R^3}{3} + \frac{(N \cdot n_c \cdot e)^2}{8\pi\varepsilon\varepsilon_0 R} - kT \cdot \log(X_N/N) \tag{1.1}$$

where $q$ is the number of contacts between the particle in a cluster, $U$ is the energy per contact, $\gamma$ is the surface tension, $R$ is the radius of the cluster, $N$ the number of the particles in a cluster, $n_c$ is the charge unit number per single particle, $e$ is the elementary charge, $\varepsilon$ and $\varepsilon_0$ are the relative dielectric constant and the permittivity in vacuum, $X_N$ is the volume fraction of molecule in the cluster. Using the following relationships between different terms the equation can be written as a function of R:

$$q = Z \cdot N = Z\left(\frac{R}{a}\right)^{d_f}; \quad U = n_E kT;$$

$$\gamma = \frac{f \cdot U \cdot N_S}{s} = f \cdot n_E kT \frac{d_f N^{(d_f-1)/d_f}}{4\pi R^2}; \quad X_N = \frac{N(4/3)\pi a^3}{(4/3)\pi R^3} = N\frac{a^3}{R^3}$$

where $Z$ is the number of particles neighbours, $a$ is the radius of a single particle, $f$ is the effective number of interactions, $d_f$ is the fractal dimension, $n_E$ the binding energy in kT units, $s$ the surface of the aggregate and $N_s$ the number of particles on the surface of an aggregate of generic fractal dimension $d_f$ [39].

Using the above conditions, we obtain equation 2 of the main text. The so obtained equation takes into account how the change from precursor to low fractality regime affects each term of the free energy. This information is enclosed in the parameter $d_f$. To perform the calculations, we use the following values:

$$kT = 4.11 \cdot 10^{-21} J; \quad a = 2 \cdot 10^{-9} m; \quad \varepsilon = 80; \quad \varepsilon_0 = 8.854 \cdot 10^{-12} C^2 J^{-1} m^{-1} \quad e = 1.602 \cdot 10^{-19} C;$$

For 2.5< df <3 we used Z=6 and f =3, for 1< df <2.5 we used Z=2 and f=1. All the calculations were performed with $n_E$=10. This value for the binding energy is used to resemble the actual reversible aggregation of proteins that includes the possibility of the rearrangement of the particle within the cluster. This is in agreement with binding energies recently estimated for aggregating ensemble of Brownian particles[40] (see also section 3 for detailed calculations). Importantly, in the specific case of aggregating proteins, the first member of the right hand side of eq. 2 in the main text refers to the binding energy between already destabilized and aggregation prone molecules.

Our coarse-grained model considers protein molecule as hard sphere interacting via electrostatic interactions. The reason for this choice is mainly dictated by the complexity of the problem and by our goal. We tried to simultaneously keep the model at a reasonable level of simplicity, whilst not neglecting some physical properties (i.e. electrostatics, energy of binding, entropy and multi-fractal nature) that are crucial in determining the large scale arrangement of the final aggregate. Including microscopic structural features of a single molecule (i.e. secondary structure) would complicate the entire approach, whilst not providing further information in terms of large scale arrangements.

**2 PBE simulations for protein aggregation: spherulites formation.**

*Model*

The simulations of protein aggregation and spherulite formation have been performed by numerically solving population balance equations (PBEs), *i.e.*, mass balances for proteins and all of their clusters. The aggregation of proteins can be modeled as a second order kinetic process. In this case PBEs are written as follows[41]:

$$\frac{dC_k}{dt} = \frac{1}{2}\sum_{i,j=1}^{i+j} K_{ij} C_i C_j - C_k \sum_{i=1}^{\infty} K_{ik} C_i , \qquad (2.1)$$

where $C_i$ is the number concentration of clusters with mass $i$ (*i.e.*, made of $i$ proteins), and $K_{ij}$ is the kernel determining the rate of aggregation between two clusters, one with mass $i$ and the other with mass $j$. The aggregation kernel used in the simulations is the conventional diffusion-limited kernel, which has the following form[42, 43]:

$$K_{ij} = \frac{2kT}{3\eta W_{ij}}(R_i + R_j)\left(\frac{1}{R_i} + \frac{1}{R_j}\right) \qquad (2.2)$$

Where $R_i$ is the outer radius of a cluster with mass $i$, $W_{ij}$ is the stability ratio of an aggregation event between the $i^{th}$ and $j^{th}$ clusters, $\eta$ the viscosity of water evaluated at the temperature of the experiments, $k$ the Boltzmann constant and $T$ the temperature.

Two types of clusters are considered in the simulations. Clusters with a radius smaller than the precursor radius $R_c$ have a fractal dimension ($d_f$) equal to 3, and their radius $R_i$ has been assumed to scale with the cluster mass as follows:

$$i = \left(\frac{R_i}{a}\right)^3 \qquad (2.3)$$

where $a$ is the protein radius. Their radius of gyration is equal to:

$$R_{g,i} = \sqrt{\frac{3}{5}} R_i \qquad (2.4)$$

Clusters with a size larger than $R_c$ have a dense precursor with a radius equal to $R_c$ and $d_f=3$, and a shell with $d_f<3$. The radial density profile of these clusters is assumed to be continuous, switching from a constant value in the precursor to a power law in the shell:

$$\rho_i(r) = \begin{cases} \dfrac{3}{4\pi a^3} & 0 < r < R_c \\ \dfrac{3}{4\pi a^3}\left(\dfrac{r}{R_c}\right)^{d_f-3} & R_c < r < R_i \end{cases} \quad (2.5)$$

The outer radius of the cluster is therefore found by integrating the density over the entire cluster volume to obtain the cluster mass:

$$i = 4\pi \int_0^{R_i} \rho_i(r) r^2 dr \quad (2.6)$$

The cluster radius of gyration is analogously found from[44]:

$$R_{g,i}^2 = \dfrac{4\pi}{i} \int_0^{R_i} \rho_i(r) r^4 dr$$

The stability ratio $W_{ij}$ in Equation 2.2 is computed from its definition[43]:

$$W_{ij} = 2 \int_2^\infty \dfrac{\exp\left(\dfrac{U_{int,ij}}{kT}\right)}{x^2} dx \quad (2.7)$$

where $U_{int,ij}$ is the interaction energy between the clusters and $x$ is the dimensionless center-to-center distance between the clusters, normalized by the average cluster radius $(R_i+R_j)/2$. The interaction energy is the sum of two contributions: an attractive Van der Waals contribution $U_{VdW,ij}$ and a repulsive electrostatic interaction term $U_{el,ij}$:

$$U_{int,ij} = U_{VdW,ij} + U_{el,ij} \quad (2.9)$$

The Van der Waals contribution has been estimated using the well-established Hamaker equation[43]:

$$U_{VdW,ij} = -\dfrac{A}{6}\left(\dfrac{2\cdot R_i \cdot R_j}{r^2-(R_i+R_j)^2} + \dfrac{2\cdot R_i \cdot R_j}{r^2-(R_i-R_j)^2} + \log\left(\dfrac{r^2-(R_i+R_j)^2}{r^2-(R_i-R_j)^2}\right)\right) \quad (2.10)$$

where $A$ is the Hamaker constant and $r$ is the (dimensional) center-to-center distance between the two clusters. For the electrostatic repulsive energy, we have used the equation proposed by Liu and Hsu[45], which is an extension for unequal surface potential of the equations proposed by Sader et al.[46]:

$$U_{el,ij} = \dfrac{\pi \varepsilon_0 \varepsilon R_i R_j}{r}\left(\dfrac{kT}{e}\right)^2 \left( \begin{array}{l} (Y_i^2+Y_j^2)\log(1-\exp(-2\kappa(r-R_i-R_j))) + \\ 4Y_i Y_j \tanh^{-1}(\exp(-\kappa(r-R_i-R_j))) \end{array} \right) \quad (2.11)$$

where $\varepsilon_0$ is the vacuum permittivity, $\varepsilon$ is water dielectric constant, $\kappa$ is the reverse Debye length, $e$ is the electron charge. The quantity $Y_i$ is defined as:

$$Y_i = 4\exp\left(\frac{\kappa}{2}(r-R_i-R_j)\right)\tanh^{-1}\left(\exp\left(-\frac{\kappa}{2}(r-R_i-R_j)\right)\tanh\left(\frac{e\varphi_i}{4kT}\right)\right) \quad (2.12),$$

and $\varphi_i$ is the surface potential of the $i^{th}$ cluster.

The above equation assumes that the surface potentials of both clusters remain constant as they approach each other. In order to find the surface potential from the surface charge density, the following equation has been used[43]:

$$\frac{\sigma_i}{kT\kappa\varepsilon_0\varepsilon} = 2\sinh\left(\frac{e\varphi_i}{2kT}\right) + \frac{4}{\kappa R_i}\tanh\left(\frac{e\varphi_i}{4kT}\right) \quad (2.13)$$

where $\sigma_i$ is the surface charge density of the $i^{th}$ cluster.

To predict the total surface charge density on one cluster, we introduce a key physically-motivated assumption depending on whether the cluster is in the precursor regime ($d_f=3$) or possesses a multi-fractal shell. In particular, for two $d_f=3$ clusters we calculate the surface charge as the one resulting from treating the clusters as dielectric spheres. This is motivated by the compact geometry which allows us to treat compact clusters as continua. As a consequence, the surface charge density for a spherical and compact cluster is given by

$$\sigma = \mathbf{P}\cdot\mathbf{n} \quad (2.14)$$

where $\mathbf{P}$ is the polarization density of the body and $\mathbf{n}$ is the unit vector normal to the surface. In a homogeneous linear and isotropic dielectric body, the polarization $\mathbf{P}$ is aligned with and proportional to the electric field $\mathbf{E}$:

$$\mathbf{P} = \varepsilon_0\chi\mathbf{E},$$

where $\varepsilon_0$ is the vacuum permittivity and $\chi$ is the electric susceptibility of the body. We can integrate both sides of Eq. (2.14) over the surface of the body, and get

$$\int\sigma dS = \varepsilon_0\chi\int\mathbf{E}\cdot\mathbf{n}\,dS$$

We can apply Gauss' law which states that

$$\int\mathbf{E}\cdot\mathbf{n}\,dS = \frac{Q}{\varepsilon_0}$$

where $Q$ is the total charge of the body (i.e. sum of the charges of each single particle within the cluster). Therefore, for an isotropic distribution of charges we have $\int\sigma dS = \sigma S$, and the surface charge density is given by

$$\sigma = \chi\frac{Q}{S} = (\varepsilon_r - 1)\frac{Q}{S}$$

where $\varepsilon_r$ is the relative permittivity of the body. For polymers, one typically has: $\varepsilon_r \sim 2-3$. Based on this, modelling our protein cluster (in the precursor regime) as a dielectric sphere, we obtain the following expression for the surface charge density for a cluster composed of $N$ protein monomers:

$$\sigma(N) = (\varepsilon_r - 1)\frac{N \times n_c e}{4\pi R^2} = (\varepsilon_r - 1)\frac{N \times n_c e}{4\pi a^2 N^{2/3}} = (\varepsilon_r - 1)\frac{N^{1/3} \times n_c e}{4\pi a^2} \qquad (2.15)$$

Where $a$ = monomer radius $R$ = cluster radius $N = \left(\dfrac{R}{a}\right)^3$ and $n_c e$ charge per monomer.

On the other hand, for clusters in the multi-fractal regime, the aggregation involves bonding between two protruding particles on the cluster outer shell which are sufficiently isolated in space from the other particles in the clusters such that the electrostatic interactions from the other particles can be safely neglected[39]. This would take into account the reduction of the electrostatic energy at decreasing $d_f$ as predicted by equation 2 in the main text. Therefore in the aggregation rate between two clusters in the multi-fractal regime we assume that the electrostatic interaction between the two clusters reduces to the interaction between two proteins protruding from the respective outer shells. The curves reported in Figure 4a in the main text have been obtained by plotting the total energy of interaction between monomer-monomer, oligomer-oligomer (close to the critical size of the precursor), shell-monomer and shell-oligomer using equation 2.10. A very short range repulsive potential has been added to prevent overlap. According to our model, all oligomers are treated as dielectric spherical particles with a surface charge density given by equation 2.15. Instead, the interactions of aggregates with a precursor-shell structure will be controlled by monomers on their outer surface. Importantly, in the simulation all the clusters, including the precursor-shell structures, can aggregate with each other. This is to reproduce with a high degree of reliability what happens in the real system. Our kinetic model considers that, after the formation of the first precursors, the growth of the shell starts, with interactions characterized by a negligible barrier (Figure 4a in the main text). In the very early stages of the shell growth, clusters (precursor-shell) are small enough to quickly diffuse in solution and the probability for these structures to interact with each other, is not negligible. This allows the formation of a small fraction of species with a shape that deviates from a perfect sphere (Figure S3a, left). Then, due to their increased size, further aggregation of such not perfectly spherical species with each other is extremely unlikely. However, they can still interact with oligomers and residual monomers in solution following the multi-fractal behaviour as predicted by the model. This leads to the formation of multi-fractal structures having a larger and not perfectly spherical central part (Figure S3a, right). It is worth noting that this is actually what experimentally happens. In Figure S3b we show that, even in a very small fraction, spherulites can clearly develop not only from a single spherical precursor but also from a more complex structure, being this in agreement with the prediction of our model (sketch in Figure S3a).

The numerical solution of the population balance equations has been performed by the Kumar-Ramkrishna method, as described elsewhere[42]. In a nutshell, a broad interval of cluster mass values is divided into intervals using a logarithmic spacing. The PBEs are solved for all the values of the boundaries of each interval, referred to as pivots. Each time an aggregation event leads to the production of an aggregate with mass value falling inside an interval, the aggregate is split between the two boundaries the interval. The splitting factors are selected so that two moments of the original distribution are conserved, specifically the zero and first moments. This procedure guaranties a high efficiency of the code, which allows one to simulate the evolution of a very broad cluster mass distribution.

*Calculation of the multi-fractal density profile and structure factor.*

To calculate the multifractal density profile (Figure 2c in the main text) the following relationship between density and $d_f$ was used:

$$\frac{\rho(R)}{\rho_{\text{Precursor}}} = R^{[d_f(R)-3]} \tag{2.16}$$

The $d_f(R)$ was extrapolated by fitting the exponential decays after precursor formation in Figure 2b and using the fitting function within equation 2.16.

The scattering structure factor of clusters with a size smaller than $R_c$ at a scattering vector $q=4\pi n/\lambda \sin(\theta/2)$, where $n$ is the refractive index of the solvent, $\lambda$ the wave length of the laser and $\theta$ the scattering angle, is computed assuming that they can be approximated as spheres[47]:

$$S_i(q) = \frac{9}{(qR_i)^6}\left(\sin(qR_i) - qR_i\cos(qR_i)\right)^2 \tag{2.17}$$

The structure factor of precursor-shell clusters is instead computed using the Fisher-Burford equation, which is commonly used to approximate the scattering behavior of fractals[43]:

$$S_i(q) = \frac{1}{\left(1 + \frac{2}{3d_f}(qR_{g,i})^2\right)^{\frac{d_f}{2}}} \tag{2.18}$$

In the case of clusters with a multi-fractal density profile $\rho(r)$ (eq. 2.16), the scattering structure factor can be computed from the following equation[47]:

$$S_i(q) = \left(4\pi \int_0^{R_i} \rho_i(r) r^2 \frac{\sin(qr)}{qr} dr\right)^2 \tag{2.19}$$

Figure S4 shows the profile (green curve) obtained from equation 2.19. The structure factor profile for the shell growth is well approximated using a constant fractal dimension of 1.9 (black dashed line). This value was used to calculate the simulated scattering curves. Curves at $d_f=1.3$ and $d_f=2.7$ are also shown for comparison.

Finally, the overall intensity of the scattered radiation by the entire cluster population at a scattering angle of $90°$ is computed through the following equation:

$$I_{90} = G\sum_{i=1}^{\infty} C_i i^2 S_i(q_{90}) \tag{2.20}$$

where $q_{90}$ is the scattering wave vector evaluated at a scattering angle of 90°, $S_i(q)$ is the scattering structure factor of a cluster with mass $i$, given by Equation (2.17) for clusters with a size smaller than $R_c$ and by Equation (2.18) for precursor-shell clusters. $G$ is a multiplicative constant depending on the experimental scattering set up, which cannot be easily determined, and therefore the scattered intensity profile height is adjusted by fitting a few experimental data points.

**3 General PBE equation includes the features of the classical nucleation theory**

The complete master kinetic equation reads as:

$$\frac{dC_k}{dt} = \frac{1}{2}\sum_{i+j=k} K_{ij} C_i C_j - C_k \sum_{i=1}^{\infty} K_{ik} C_i - K_k^B C_k + \sum_{i=k+1}^{\infty} K_{ik}^B C_i \qquad (3.1)$$

where the last two terms accounts for thermal breakup of a cluster of size $k$ and generation of a $k$ cluster by breakup of a cluster of size $k+i$. These terms are required when the inter-protein attraction is such that thermal energy can cause the complete dissociation of the bond between two proteins on a time scale comparable to the diffusive attachment time scale of a monomer. In our model the binding energy between particles is ~10kT, such that the rate of detachment of a monomer from a cluster (assuming it is bound to two particles on the surface) is according to a previous report[40], $\sim (D/\delta^2)\exp[-2\cdot(10/kT)] \sim 2\cdot 10^{-3}\,\mathrm{s}^{-1}$, where we used the typical range of hydrophobic attraction which is of the order of 10 nm and rather independent of the chemical composition of the approaching surfaces[16, 40]. For the association rate of a monomer under diffusion-limited conditions we have $\sim (8/3)(kT/\eta)(N/V) \sim 4.8\,\mathrm{s}^{-1}$, from the standard Smoluchowski rate. Clearly, in our system the mismatch between association and dissociation rates is significant enough to neglect the last two terms in Eq. (3.1) leading to the use of Eq. 4 of the main text. Importantly, the absolute value that we find for the dissociation rate is such that particles certainly can rearrange during the aggregation process allowing us for setting up an equilibrium-like free energy of clustering.

However, in the general case of reactions where thermal breakup is significant, all the terms in Eq (3.1) should be considered. If we were really close to the metastability region of the proteins in water (i.e. close to the binodal line for equilibrium liquid-liquid phase separation), then a nucleation scenario within this approach can be recovered. In that case detailed balance is exactly satisfied and clusters are formed due to critical fluctuations under supersaturation conditions. Under these conditions, clusters are highly localized and noninteracting and grow very slowly by means of one-step particle attachment. Therefore, only terms of the type $K_{k-1,1}C_{k-1}$ and $K_{k,1}C_k$ contribute to the first and the second term, respectively, on the r.h.s. of Eq. (3.1). Note that we have incorporated the monomer concentration $C_1$ in the rate constants. Under conditions of localized fluctuational growth, Eq.(3.1) in the initial stage of aggregation reduces to:

$$\frac{dC_k}{dt} = K_{k-1,1}C_{k-1} - K_{k,1}C_k - K_k^B C_k + K_{k+1,1}^B C_{k+1}$$

To shorten the notation we put $K_{k-1,1} \equiv K_{k-1}$ etc. and rewrite the equation as:

$$\frac{dC_k}{dt} = K_{k-1}C_{k-1} - K_k C_k - K_k^B C_k + K_{k+1}^B C_{k+1} \qquad (3.2)$$

Since the attraction is weak and thermal dissociation is important, the principle of detailed balance is applicable in this limit. Hence, we now introduce the equilibrium or steady-state concentration of aggregates of size k as $C_k^{eq}$ which is a Boltzmann function of the minimum work $\Delta F$ (free energy) needed to form an aggregate of size $k$: $C_k^{eq} \sim \exp(-\Delta F/kT)$. Upon applying the principle of detailed balance we have:

$$C_k^{eq} K_k = C_{k+1}^{eq} K_{k+1}^{B}$$
$$C_{k-1}^{eq} K_{k-1} = C_k^{eq} K_k^{B}$$
(3.3)

These relations allow us to eliminate from Eq. (3.2) the quantities $K_k^B$ and $K_{k+1}^B$, and Eq. (3.2) becomes:

$$\frac{dC_k}{dt} = K_k \left[ -C_k + C_{k+1} C_k^{eq} / C_{k+1}^{eq} \right] + K_{k-1} \left[ C_{k-1} - C_k C_{k-1}^{eq} / C_k^{eq} \right]$$
$$= K_k C_k^{eq} \left[ C_{k+1} / C_{k+1}^{eq} - C_k / C_k^{eq} \right] - K_{k-1} C_{k-1}^{eq} \left[ C_k / C_k^{eq} - C_{k-1} / C_{k-1}^{eq} \right]$$
(3.4)

Let us now transform the discrete distribution $C_k$ (discrete in the cluster size $k$) into a continuous one $C(x)$ where $x$ is a continuous variable expressing the cluster size. Denoting by $\lambda$ the spacing along the $x$ axis between the neighbouring sizes $k$ and $k+1$, we have $C_k = \lambda C(x)$, $C_{k+1} = \lambda C(x+\lambda)$ etc. Since $\lambda$ is constant and $C$, $C^{eq}$, and $K$ vary little within the length $\lambda$, one can do an expansion in power series of $\lambda$ and retain only the first non-vanishing term. Using this procedure, Zeldovich[48] has shown that Eq.(3.4) reduces to the following form:

$$\frac{\partial C}{\partial t} = -\frac{\partial J}{\partial x}$$
(3.5)

Where $J$ is the flux in "size space" given by: $J = -D \partial C / \partial x + A \cdot C$. Here $D = \lambda K$ is the diffusion coefficient in "size space" while $A = -D \cdot \Delta F'(x) / kT$ is the drift coefficient. Therefore, our original equation can be reduced, under the assumptions stated above, to a diffusion equation (in size space) in the field of force of the free energy of aggregation. Under the circumstances that the attraction is weak in comparison with the surface energy of the cluster (i.e. the case for aggregation reactions) the free energy might go through a local maximum at a size $x^*$ due to the competition between attraction and surface energy. Hence, clusters smaller than $x^*$ tend to shrink whereas clusters $> x^*$ tend to grow. Then $\Delta F$ can be expanded to second order near the maximum: $F = F(x^*) - \frac{1}{2} F''(x^*)(x - x^*)^2$, and Eq. (3.5) can be solved analytically at the steady-state ($\partial C / \partial t = 0$) in the standard way of Kramers by means of the saddle-point method[49]. This yields the well-known formula of nucleation theory for the nucleation rate[50, 51]:

$$J = C_1 K_{k^*,1} \left( \frac{\Delta F''(x^*)}{2\pi kT} \right) \exp\left( -\frac{\Delta F(x^*)}{kT} \right)$$

Importantly, in the case of spherulites a process different from the standard nucleation/elongation determines the growth of the aggregates, i.e. the multi-fractal growth, and, as a consequence, no use of the rate J can be found. Finally, it is important to note that this derivation cannot be applied to earlier models because the reduction of Eq.(3.1) to Eq.(3.2) requires that one takes the interactions into account in the physical formulation of the microscopic rates $K_{ij}$. If the rates are taken as fitting parameters, it is impossible to recover nucleation theory.

**Supplementary Figures and Legends**

**Supplementary Figure S1**

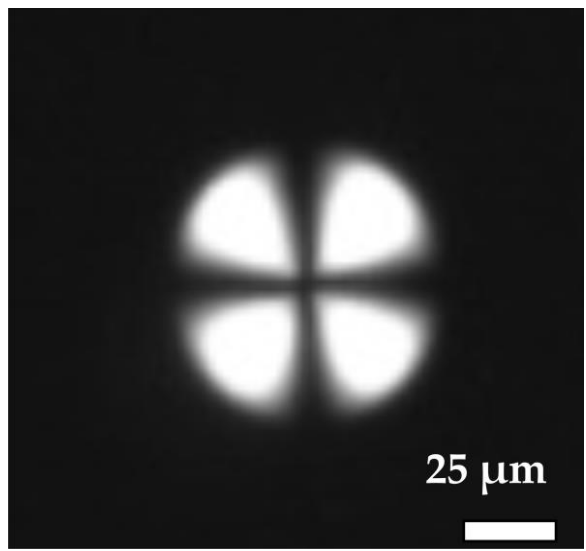

**Supplementary Figure S1: Optical microscopy on amyloid spherulites.** Amyloid spherulites as they appear in solution under crossed polarized optical microscope

**Supplementary Figure S2**

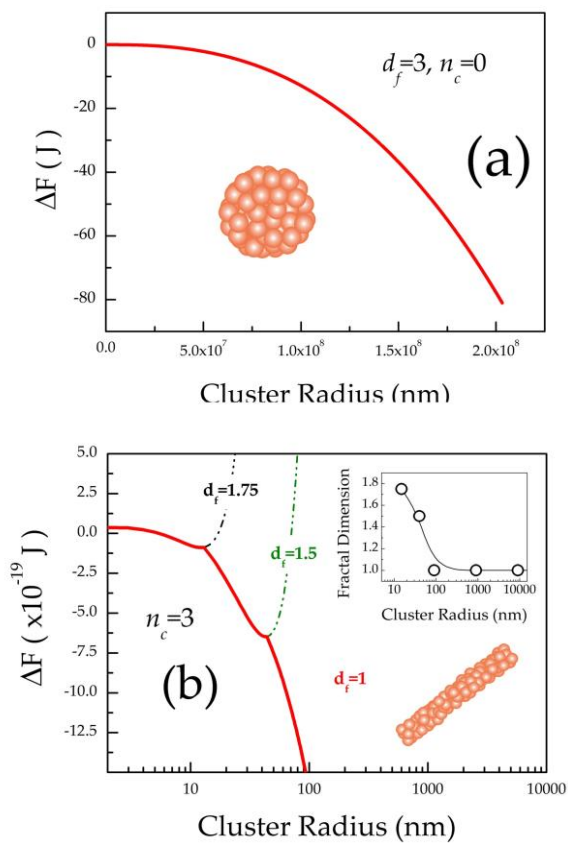

**Supplementary Figure S2: Morphologies as a function of the protein charge.** Free energy landscape for a growing aggregate as a function of the aggregate radius calculated by means of equation 2 (see main text) at (a) $d_f=3$ and $n_c=0$ and (b) at different fractal dimensions and with $n_c=3$. In the case of $n_c=0$, the growth proceeds with $d_f=3$ for aggregates with radii up to several cm. This means that for the range of sizes of protein aggregates experimentally observed, the growth will basically proceed as a sphere with a constant density. On the other hand, at $n_c=3$, the aggregate grows with $1<d_f<2$ since the very early stages of the process and then the growth proceeds with $d_f=1$ up to several cm (inset in Figure S2b). This means that a two-step process takes place: an early formation of a species with df slightly higher than 1 and then a linear growth of the aggregate until the reaction reaches its completion. This can resemble the classical description based on the nucleation and elongation proposed for simple elongated fibrils

**Supplementary Figure S3**

(a)
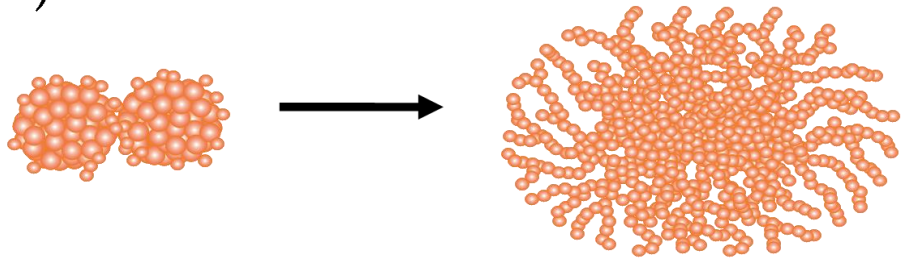

(b)
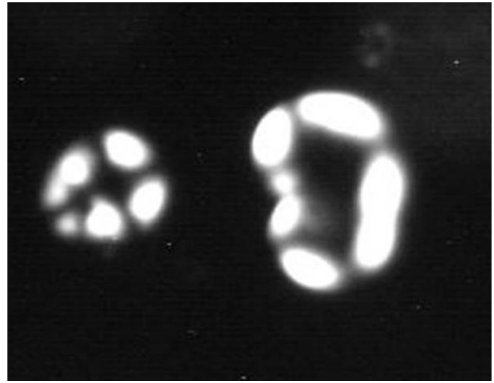

**Supplementary Figure S3: Spherulites growing from complex precursor.** (a) Sketch of the mechanism bringing to the formation of a small fraction of multi-fractal structures on a complex precursor. (b) Amyloid spherulites with a shell developing from a complex precursor as experimentally observed. The shell grows on two collapsed central parts.

**Supplementary Figure S4**

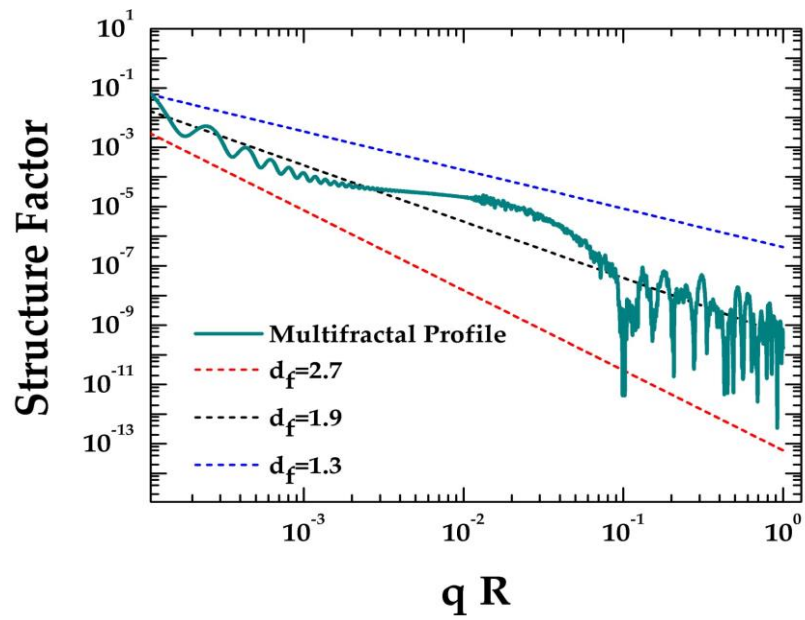

**Supplementary Figure S4: Structure factor in multifractal regime.** Structure factor as calculated from equation 2.18 (red, black and blue dashed lines) and equation 2.19 (green solid line) for fractal dimensions 2.7, 1.9, 1.3 and for the multi-fractal density, respectively.

**Supplementary Tab S1**

| pH | Effective Charge | Precursor Radius (nm) |
|---|---|---|
| 1 | 0.599 | 31.6367 |
| 1.25 | 0.597 | 31.7491 |
| 1.5 | 0.582 | 32.6178 |
| 1.75 | 0.555 | 34.3036 |

**Supplementary Table S1:** Pairs of precursor radius and charge values used for the simulations of the spherulites kinetics as a function of the pH.